\documentclass[aps,preprint,nofootinbib]{revtex4}
\usepackage{amsmath}
\usepackage{amssymb}
\usepackage{graphicx}
\usepackage{xcolor}
\usepackage{slashed}
\usepackage{mathtools}
\usepackage{capt-of}
\usepackage[hypertexnames=false,hyperfootnotes=false]{hyperref}
\usepackage{placeins}
\usepackage{flafter}
\usepackage{tabularx}
\usepackage{booktabs}
\usepackage{multirow}
\graphicspath{{figures/}}

\allowdisplaybreaks[4]

\newcommand{\Dm}{\mathcal D_\mu}
\newcommand{\eps}{\epsilon}
\newcommand{\as}{\alpha_s}

\newcommand{\dd}{\mathrm{d}}
\newcommand{\order}{\mathcal O}
\newcommand{\GeV}{\mathrm{GeV}}
\newcommand{\fb}{\mathrm{fb}}
\newcommand{\Hcal}{\mathcal H}
\newcommand{\Ecal}{\mathcal E}
\newcommand{\Htcal}{\widetilde{\mathcal H}}
\newcommand{\Etcal}{\widetilde{\mathcal E}}
\newcommand{\Gcal}{\mathcal G}

\begin{document}

\title{Exclusive Leptonium Electroproduction\\[0.6cm]}

\author{\vspace{0.3cm} Hao-ye Deng$^{1,3}$, Qi-Ming Feng$^{1}$\footnote[1]{fengqiming@ucas.ac.cn}, Qi-Wei Hu$^{1,2}$, Si-Qin Huang$^{1,3}$, Cong-Feng Qiao$^{1,3}$\footnote[2]{qiaocf@ucas.ac.cn}, Jia-Xuan Shen$^{1,2}$, Ting-Ting Wang$^{1,3}$, Shun-Yan Yu$^{1,2}$, Hao Zhang$^{2,1,4}$\footnote[3]{zhanghao@ihep.ac.cn}, Xuan-Heng Zhang$^{1}$, and Yi-Nan Zhao$^{1,2}$\vspace{0.3cm}}

\affiliation{\small {$^1$ School of Physics, University of Chinese Academy of
Sciences, Beijing 100049, China\\
$^2$ Institute of High Energy Physics, Chinese Academy of Sciences, Yuquan Road 19B, Beijing 100049, China\\
$^3$ International Centre for Theoretical Physics Asia-Pacific, Beijing 100190, China\\
$^4$ Center for High Energy Physics, Peking University, Beijing 100871, China}
}

\author{~\\}

\begin{abstract}
Purely leptonic bound states provide precision probes of QED.  Positronium
$(e^+e^-)$ and muonium $(\mu^+e^-)$ have long been observed, whereas
dimuonium $(\mu^+\mu^-)$ and tauonium $(\tau^+\tau^-)$ remain undiscovered.
We study exclusive vector-leptonium electroproduction in $ep$ collisions
within nonrelativistic QED.  We include the Bethe--Heitler and double deeply
virtual Compton scattering contributions and their interference, and calculate
the NLO QCD hard-scattering kernels entering the dominant Compton form factor
$\Hcal$ within collinear GPD factorization.  The NLO QCD correction
to the DDVCS contribution changes from a strong suppression at low photon
virtuality to a sizable enhancement as the lower virtuality cut is raised,
with the gluon channel providing the dominant contribution.  Bethe--Heitler production dominates the exclusive rate, supporting
dedicated dimuonium searches at the EIC and JLab, with larger samples expected
at higher-energy electron--proton colliders.  The much larger positronium
samples provide a high-statistics environment for precision QED studies,
whereas tauonium production remains strongly suppressed.
\end{abstract}
\maketitle
\clearpage
\raggedbottom

\section{Introduction}
\label{sec:introduction}

Purely leptonic bound states provide precision probes of QED.  Positronium $(e^+e^-)$ was discovered in 1951 and muonium $(\mu^+e^-)$ in 1960~\cite{Deutsch:1951ps,Hughes:1960mu}, whereas the heavier dimuonium $(\mu^+\mu^-)$ and tauonium $(\tau^+\tau^-)$ states remain unobserved.  
Dimuonium is considerably more compact than positronium and correspondingly more sensitive to short-distance effects~\cite{Jentschura:1997tv}.
Its vector ground state is particularly relevant for production studies because it couples directly to the electromagnetic current and has a prominent $e^+e^-$ decay mode.  
These features have motivated a broad range of theoretical and experimental proposals for producing and identifying true leptonium under complementary laboratory conditions~\cite{Brodsky:2009gx,Banburski:2012tk,Fox:2022bpc,Feng:2025dm,Feng:2026upc,Serri:2026nrqed,Francener:2024qedbound,CidVidal:2019qub,Bertulani:2023nch,Martynenko:2024rfj,Gargiulo:2025pmu,Zhao:2025gia}.

Lepton--hadron collisions offer a complementary production environment.
Previous studies addressed inclusive or proton-dissociative leptonium
production with automated calculations in nonrelativistic QED (NRQED) and
coherent two-photon production in electron--ion collisions
~\cite{Serri:2026nrqed,Francener:2024qedbound}.
In the exclusive reaction considered here, the recoil proton remains intact and the hadronic amplitude probes off-forward light-cone matrix elements. 
The resulting generalized parton distributions (GPDs) reduce to ordinary parton distributions in the forward limit and have moments related to elastic form factors~\cite{Ji:1996ek,Radyushkin:1997ki,Collins:1998be,Diehl:2003ny}. 
Their convolution with perturbative coefficient functions defines the Compton form factors (CFFs) entering hard exclusive amplitudes. 
HERA established deeply virtual Compton scattering (DVCS) and exclusive vector-meson production as measurable collider processes~\cite{H1:2007vrx,H1:2009wnw}.
High-luminosity lepton--hadron programs at JLab and the EIC, together with the proposed EicC and LHeC facilities, extend both the luminosity and kinematic reach of such studies~\cite{Baltzell:2020qoo,AbdulKhalek:2022eic,Anderle:2021eicc,Agostini:2021lhec}.

Exclusive vector-dimuonium electroproduction proceeds through the Bethe--Heitler (BH) and double
deeply virtual Compton scattering (DDVCS) amplitudes.  The BH amplitude is
determined by the elastic proton form factors, whereas the DDVCS amplitude
contains the GPD-dependent CFFs; their interference is linear in the CFFs
~\cite{Belitsky:2003fj,Deja:2023ahc}.  In continuum DDVCS,
the spacelike and timelike photon virtualities can be varied independently,
providing access to GPDs away from the DVCS crossover line
~\cite{Guidal:2002kt}.  For dimuonium, the outgoing timelike
virtuality is fixed by the bound-state mass, so the subprocess follows a
fixed-mass trajectory and approaches the DVCS limit as the incoming
spacelike virtuality increases.

The one-loop DVCS coefficient functions and subsequent NLO studies show that the
singlet gluon term can be sizable and can cancel the quark contribution at small
skewness~\cite{Ji:1998xh,Freund:2001dvcsobs,Moutarde:2013nlo}; this cancellation remains important at
NNLO~\cite{Braun:2022nnlo}.  General DDVCS coefficient functions are known at one
and two loops~\cite{Pire:2011st,Braun:2024jns}, while exclusive vector-meson
calculations illustrate the associated scale sensitivity
~\cite{Ivanov:2004ax,Chen:2019pec,Flett:2021ghh}.

In this work, we study exclusive vector-leptonium electroproduction in
$ep$ collisions through the BH and DDVCS mechanisms, retaining their
interference and calculating the dominant charge-even vector-$\Hcal$ channel
at NLO in QCD.  We compare representative CFF inputs, examine the photon-virtuality dependence,
and evaluate the Born BH production rates across the vector-leptonium family at
representative facilities.
The remainder of the paper is organized as follows. 
Section~\ref{sec:formalism} presents the formalism and details of our calculation.
Section~\ref{sec:numerical-results} describes the numerical setup and results. 
Section~\ref{sec:summary} summarizes our results and gives out some outlooks.

\section{Formalism and Calculation}
\label{sec:formalism}

\subsection{LO amplitudes}

\begin{figure}[!htbp]
\centering
\includegraphics[width=0.98\textwidth]{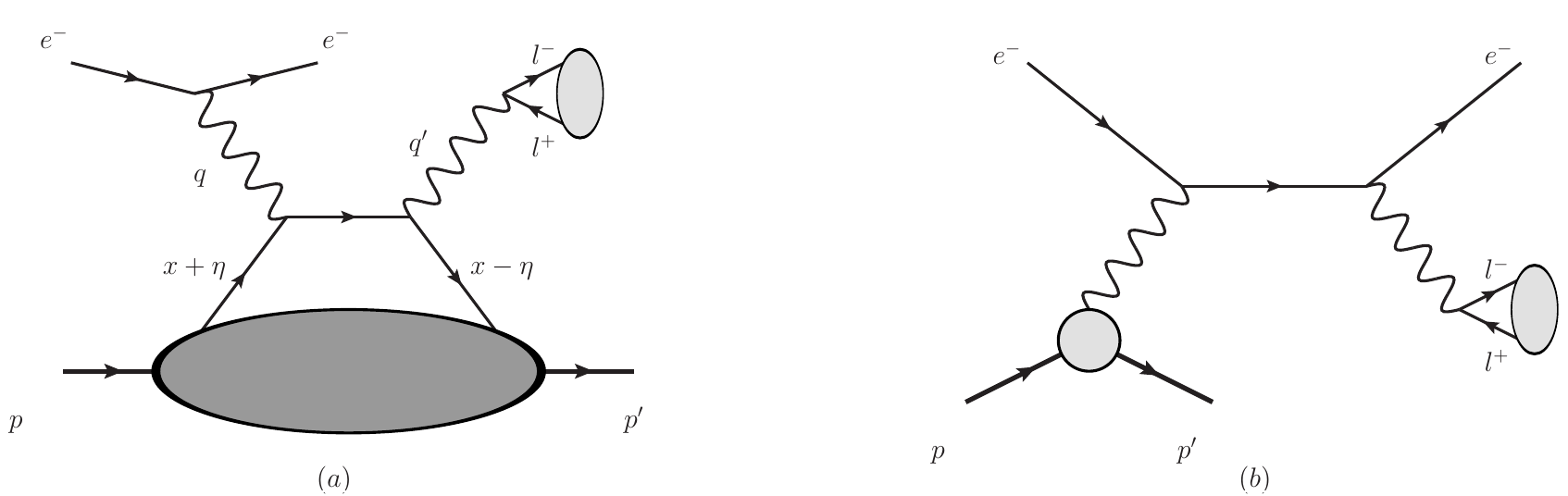}
\caption{Representative Born-level mechanisms for exclusive
vector-dimuonium electroproduction.  Panel (a) shows the DDVCS handbag
subprocess, with the outgoing timelike photon projected onto the ${}^3S_1$
bound state; panel (b) shows one of the two electron-line BH orderings.}
\label{fig:production-mechanisms}
\end{figure}

Exclusive vector-dimuonium electroproduction proceeds through the DDVCS and BH mechanisms shown in Fig.~\ref{fig:production-mechanisms}.
The corresponding process is
\begin{equation}
 e(k)+p(p)\to e(k')+p(p')+\Dm(P,\lambda_V)\; ,
 \label{eq:process}
\end{equation}
 with $\Dm =(\mu^+\mu^-)({}^3S_1)$.  The DDVCS and electron-line BH contributions have a common form for all
vector-leptonium states; only the lepton mass and the wave function at the
origin are state dependent.  At LO, the two amplitudes contribute to the
same final state. 
We therefore define $\mathcal M^{(0)}=\mathcal M_{\mathrm{DDVCS}}^{(0)}+\mathcal M_{\mathrm{BH}}^{(0)}$ and write the physical cross section as
\begin{align}
 \dd\sigma_{\mathrm{LO}}
 &=\frac{1}{2\sqrt{\lambda(s,m_e^2,m_p^2)}}\,
 \overline{|\mathcal M^{(0)}|^2}\,\dd\Phi_3
 \nonumber\\
 &=\frac{1}{2\sqrt{\lambda(s,m_e^2,m_p^2)}}\,
 \overline{|\mathcal M_{\mathrm{DDVCS}}^{(0)}+
 \mathcal M_{\mathrm{BH}}^{(0)}|^2}\,\dd\Phi_3\ .
 \label{eq:cross-section}
\end{align}
Here, $s=(k+p)^2$ and $\lambda(a,b,c)=a^2+b^2+c^2-2ab-2ac-2bc$ is the K\"all\'en function.  The bar
denotes the average over the initial spins and the sum over the final spins and
vector-state polarizations, and $\dd\Phi_3$ is the Lorentz-invariant three-body
phase-space measure.

We begin with the DDVCS contribution.  Its hadronic subprocess is
$\gamma^*(q)p(p)\to\gamma^*(P)p(p')$, and the outgoing timelike
electromagnetic current is matched onto the vector-dimuonium NRQED matrix
element.  The relevant invariants are
\begin{equation}
 \begin{aligned}
 q&=k-k', & \Delta&=p'-p=q-P,\\
 Q^2&=-q^2, & W^2&=(p+q)^2,\\
 t&=\Delta^2, & P^2&=M_{\Dm}^2.
 \end{aligned}
 \label{eq:compton-kinematics}
\end{equation}
The outgoing virtuality is fixed by the dimuonium mass rather than scanned as
 an independent DDVCS variable.  Consequently, the scaling variables defined
 in Eq.~\eqref{eq:xi-rho} behave as $\rho\to\xi$, and the
subprocess approaches the DVCS limit when $Q^2\gg M_{\Dm}^2$.

 NRQED separates the short-distance production of the nearly
 nonrelativistic $\mu^+\mu^-$ pair from its Coulombic binding.  At LO in the
 relative-velocity expansion, the vector state is represented by the
spin-triplet $S$-wave matching matrix~\cite{Feng:2025dm},
\begin{equation}
 \Pi_3(P,\lambda_V)=-\frac{f_{\Dm}}{4}
 \slashed\eps^{\,*}(P,\lambda_V)(\slashed P+M_{\Dm}).
 \label{eq:bound-state}
\end{equation}
At this accuracy, $M_{\Dm}=2m_\mu$,
$f_{\Dm}^2=4|\psi_{1S}(0)|^2/M_{\Dm}$, and
$|\psi_{1S}(0)|^2=(\alpha m_\mu)^3/(8\pi)$.  Separating the electron current
from the DDVCS tensor gives
\begin{equation}
 \mathcal M_{\mathrm{DDVCS}}^{(0)}=\frac{e^2}{Q^2}
 \bar u(k')\gamma_\nu u(k)\,\eps_\mu^*(P,\lambda_V)
 T^{(0)\mu\nu}.
 \label{eq:compton-master-amplitude}
\end{equation}

In the generalized Bjorken limit, $T^{(0)\mu\nu}$ factorizes into
short-distance coefficients and twist-2 GPDs.  Following
Ref.~\cite{Deja:2023ahc}, the twist-2 hard tensor is evaluated for an
auxiliary collinear event with the same $Q^2$, $W$, and $M_{\Dm}$ but with
$t=t_0\equiv t_{\min}$.  Denoting the corresponding final proton and
dimuonium momenta by $p'_0$ and $P_0$, we define
$\Delta_0=p'_0-p$, $\bar p_0=(p+p'_0)/2$, and
$\bar q_0=(q+P_0)/2$.  The associated scaling variables are
\begin{equation}
 \xi=-\frac{\Delta_0\cdot\bar q_0}{2\bar p_0\cdot\bar q_0},
 \qquad
 \rho=-\frac{\bar q_0^2}{2\bar p_0\cdot\bar q_0}
 =\xi\frac{Q^2-M_{\Dm}^2+t_0/2}{Q^2+M_{\Dm}^2}.
 \label{eq:xi-rho}
\end{equation}
The auxiliary event fixes the scaling variables, light-cone basis, and
hard-tensor structures, whereas the external electron current and dimuonium
polarization vector use the physical momenta.  The physical value of $t$ is
retained in the CFFs and phase space, and the BH amplitude is evaluated
entirely with physical momenta.  Differences associated with $P-P_0$ are
finite-$t$ kinematic corrections beyond the present twist-2 accuracy.

The light-cone basis is fixed directly by the auxiliary momenta.  Defining
\begin{equation}
 \begin{aligned}
 R_0&=\sqrt{(\bar p_0\!\cdot\!\bar q_0)^2-\bar p_0^2\bar q_0^2},
 &\alpha_0&=\frac{\bar q_0^2}{\bar p_0\!\cdot\!\bar q_0+R_0},\\
 n_{\mathrm{LC}}^\mu&=\frac{\bar q_0^\mu-\alpha_0\bar p_0^\mu}{R_0},
 &p_{\mathrm{LC}}^{\mu}&=\bar p_0^\mu-\frac{\bar p_0^2}{2}n_{\mathrm{LC}}^\mu,
 \end{aligned}
 \label{eq:light-cone-basis}
\end{equation}
we have $n_{\mathrm{LC}}^2=p_{\mathrm{LC}}^2=0$,
$p_{\mathrm{LC}}\cdot n_{\mathrm{LC}}=1$, and
$\bar p_0\cdot n_{\mathrm{LC}}=1$.  The transverse tensors are
\begin{equation}
 g_\perp^{\mu\nu}=g^{\mu\nu}
 -p_{\mathrm{LC}}^\mu n_{\mathrm{LC}}^\nu
 -n_{\mathrm{LC}}^\mu p_{\mathrm{LC}}^\nu,
 \qquad
 \eps_\perp^{\mu\nu}=\eps^{\mu\nu\alpha\beta}
 p_{\mathrm{LC},\alpha}n_{\mathrm{LC},\beta}.
 \label{eq:transverse-tensors}
\end{equation}
The factorized LO DDVCS tensor is
\begin{equation}
 T^{(0)\mu\nu}=\frac{e^2f_{\Dm}}{2M_{\Dm}}
 \left(g_\perp^{\mu\nu}\mathcal J_V-i\eps_\perp^{\mu\nu}\mathcal J_A\right).
 \label{eq:compton-amplitude}
\end{equation}
The nonforward proton matrix elements are parametrized by the four twist-2
GPDs $H^q$, $E^q$, $\widetilde H^q$, and $\widetilde E^q$.  After
convolution with the direct and crossed coefficient functions, they enter
through
\begin{align}
 \mathcal J_V={}&\bar u(p')\left[
 \slashed{n}_{\mathrm{LC}}\,\Hcal
 +\frac{i\sigma^{\alpha\beta}n_{\mathrm{LC},\alpha}\Delta_\beta}{2m_p}\Ecal
 \right]u(p),
 \nonumber\\
 \mathcal J_A={}&\bar u(p')\left[
 \slashed{n}_{\mathrm{LC}}\,\Htcal
 +\frac{\Delta\cdot n_{\mathrm{LC}}}{2m_p}\Etcal
 \right]\gamma_5u(p).
 \label{eq:compton-currents}
\end{align}
For quark charges $e_q$ in units of $e$, the CFFs are
\begin{align}
 \{\Hcal,\Ecal\}(\rho,\xi,t;\mu_F)&=
 \sum_q e_q^2\int_{-1}^{1}\dd x\,
 \left(\frac{1}{x-\rho+i0}+\frac{1}{x+\rho-i0}\right)
 \{H^q,E^q\}(x,\xi,t;\mu_F),
 \nonumber\\
 \{\Htcal,\Etcal\}(\rho,\xi,t;\mu_F)&=
 \sum_q e_q^2\int_{-1}^{1}\dd x\,
 \left(\frac{1}{x-\rho+i0}-\frac{1}{x+\rho-i0}\right)
 \{\widetilde H^q,\widetilde E^q\}(x,\xi,t;\mu_F).
 \label{eq:cff-definitions}
\end{align}
All numerical CFF inputs are converted to this convention before they are
contracted with Eq.~\eqref{eq:compton-amplitude}.
The convolutions cover the Dokshitzer--Gribov--Lipatov--Altarelli--Parisi
(DGLAP) and Efremov--Radyushkin--Brodsky--Lepage (ERBL) regions.  The $i0$ prescription
fixes the imaginary part, while the principal-value integral determines the
real part.

The BH contribution is determined by the elastic proton current,
\begin{equation}
 J_p^\alpha=\bar u(p')\left[F_1(t)\gamma^\alpha
 +\frac{i\sigma^{\alpha\beta}\Delta_\beta}{2m_p}F_2(t)\right]u(p),
 \label{eq:proton-current}
\end{equation}
where $F_1$ and $F_2$ are the Dirac and Pauli form factors.  Charge
conjugation of the neutral $C=-1$ state eliminates the subamplitudes in which
the proton-exchanged photon couples to the produced muon line.  The two
surviving electron-line orderings give
\begin{align}
 \mathcal M_{\mathrm{BH}}^{(0)}={}&\frac{e^4f_{\Dm}}{tM_{\Dm}}J_p^\alpha\,
 \bar u(k')\left[
 \gamma_\alpha\frac{\slashed k-\slashed P+m_e}{D_-}\gamma_\mu
 +\gamma_\mu\frac{\slashed k'+\slashed P+m_e}{D_+}\gamma_\alpha
 \right]u(k)\,\eps^{*\mu}(P,\lambda_V)\ ,
 \label{eq:bh-amplitude}
\end{align}
with $D_- = (k-P)^2-m_e^2 + i0$ and $D_+ = (k'+P)^2-m_e^2+i0$. The BH amplitude retains the physical external momenta, the electron mass, and the elastic form factors without a collinear expansion.  

\subsection{NLO QCD correction}
\label{sec:nlo}

The NLO QCD calculation is performed for the amplitude associated with $\Hcal$
and includes the diagonal-quark and gluon coefficient functions,
\begin{equation}
 \mathcal M_{\mathrm{DDVCS}}=\mathcal M_{\mathrm{DDVCS}}^{(0)}
 +\mathcal M_q^{(1)}+\mathcal M_g^{(1)}+\order(\alpha_s^2).
 \label{eq:nlo-amplitude-expansion}
\end{equation}
Writing $\mathcal M^{(1)}=\mathcal M_q^{(1)}+\mathcal M_g^{(1)}$, the NLO and NLO$^{*}$ cross sections are defined as
\begin{align}
 \dd\sigma_{\mathrm{NLO}}
 &=\frac{1}{2\sqrt{\lambda(s,m_e^2,m_p^2)}}
 \left[\overline{|\mathcal M^{(0)}|^2}
 +2\operatorname{Re}\overline{\mathcal M^{(0)*}\mathcal M^{(1)}}\right]
 \dd\Phi_3,
 \label{eq:fixed-order}\\
 \dd\sigma_{\mathrm{NLO}^{*}}
 &=\frac{1}{2\sqrt{\lambda(s,m_e^2,m_p^2)}}
 \overline{|\mathcal M^{(0)}+\mathcal M^{(1)}|^2}\,\dd\Phi_3
 \nonumber\\*
 &=\dd\sigma_{\mathrm{NLO}}
 +\frac{1}{2\sqrt{\lambda(s,m_e^2,m_p^2)}}
 \overline{|\mathcal M^{(1)}|^2}\,\dd\Phi_3.
 \label{eq:nlo-star-definition}
\end{align}
For a DDVCS-only prediction, $\mathcal M^{(0)}$ is replaced by
 $\mathcal M_{\mathrm{DDVCS}}^{(0)}$.  For comparison with the amplitude-level
prescriptions used in Refs.~\cite{Ivanov:2004ax,Chen:2019pec,Flett:2021ghh},
we also evaluate the square of the amplitude truncated at one loop.  Since it
contains $|\mathcal M^{(1)}|^2$, NLO$^{*}$ is quoted separately.  The BH amplitude is unchanged by the
QCD hard correction, which enters the total rate through DDVCS and its
interference with BH.

\begin{figure}[!htbp]
\centering
\includegraphics[width=0.92\textwidth]{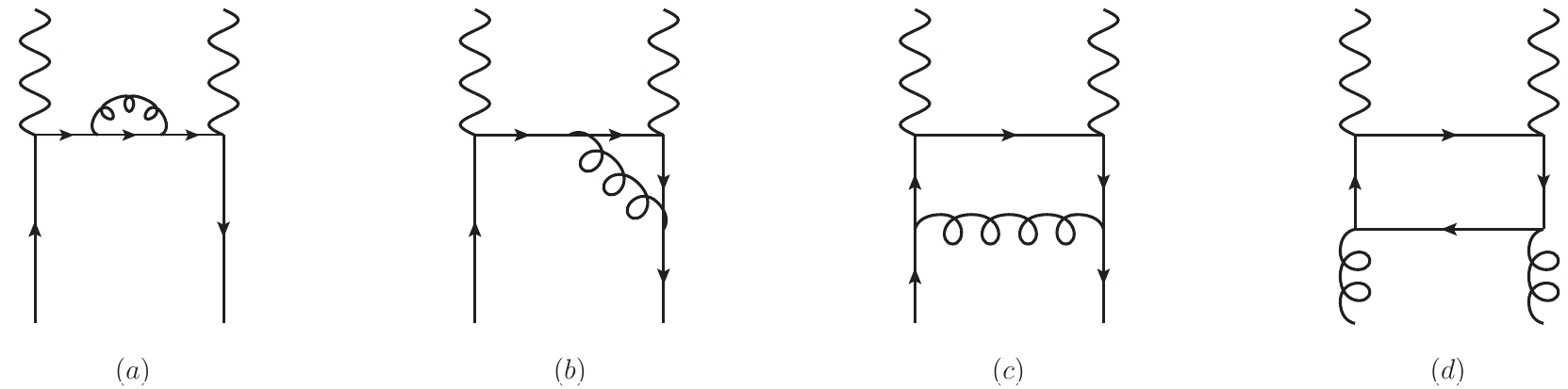}
\caption{Representative one-loop topologies contributing to the partonic
hard-scattering kernel.  Panels (a)--(c) illustrate the self-energy, vertex,
and box classes of the eight nonvanishing diagonal-quark diagrams; panel (d)
illustrates one of the six quark-box diagrams matched onto the gluon GPD.
Crossed orderings and counterterm insertions are not shown.}
\label{fig:partonic-hard}
\end{figure}

After the ${}^3S_1$ projection and symmetry relations are imposed, the
diagonal-quark channel contains eight nonvanishing one-loop diagrams, grouped
into the self-energy, vertex, and box classes shown in
Figs.~\ref{fig:partonic-hard}(a)--(c).  The gluon channel starts at this order
with six quark-box diagrams matched onto two collinear external gluons, as
represented by Fig.~\ref{fig:partonic-hard}(d).  Crossed orderings and
counterterm insertions are included in the calculation.
Dimensional regularization with $D=4-2\varepsilon$ is employed throughout this calculation.  The
massless external light-quark field is renormalized in the on-shell scheme;
its scaleless self-energy contains ultraviolet and infrared poles with
opposite signs.  Keeping the two pole labels distinct, the on-shell field
counterterm removes the ultraviolet pole, while the remaining collinear
infrared pole is absorbed into the singlet GPD counterterm.  The diagrams and amplitudes are generated with
\textsc{FeynArts}~\cite{Hahn:2000kx}, the Dirac, Lorentz, color, and bound-state algebra is
performed with \textsc{FeynCalc}~\cite{Shtabovenko:2020gxv}, and the scalar loop integrals are reduced
with \textsc{FIRE6}~\cite{Smirnov:2019qkx}.  The finite kernels are exported to the C++ event code,
and the phase-space integrations are performed with \textsc{CUBA}~\cite{Hahn:2004fe}.

At this order, the required operator redefinition is the first row of the
singlet evolution matrix,
\begin{align}
 F^\Sigma(x,\xi,t;\mu_F)={}&F^\Sigma_{\mathrm{bare}}(x,\xi,t)
 -R_\varepsilon\Big[V_{qq}^{[0]}\otimes F^\Sigma_{\mathrm{bare}}
 +V_{qg}^{[0]}\otimes F^g_{\mathrm{bare}}\Big](x,\xi,t),
 \nonumber\\
 R_\varepsilon={}&\frac{\as}{2\pi\varepsilon}
 \frac{\Gamma(1-\varepsilon)}{\Gamma(1-2\varepsilon)}
 \left(\frac{4\pi\mu_R^2}{\mu_F^2}\right)^\varepsilon.
 \label{eq:singlet-gpd-renormalization}
\end{align}
After the common electromagnetic and bound-state normalization is factored
out, the finite quark and gluon tensors are expanded in the same five-element
basis,
\begin{align}
 T_{r,\mathrm{red,fin}}^{(1)\mu\nu}
 &=\sum_{i\in\{T,1,\ldots,4\}}C_i^r\tau_i^{\mu\nu},
 \qquad r=q,g,
 \nonumber\\
 (\tau_T,\tau_1,\tau_2,\tau_3,\tau_4)^{\mu\nu}
 &=\left(g_\perp^{\mu\nu},n_{\mathrm{LC}}^\mu n_{\mathrm{LC}}^\nu,
 p_{\mathrm{LC}}^\mu n_{\mathrm{LC}}^\nu,
 n_{\mathrm{LC}}^\mu p_{\mathrm{LC}}^\nu,
 p_{\mathrm{LC}}^\mu p_{\mathrm{LC}}^\nu\right).
 \label{eq:nlo-factorization}
\end{align}
The collinear poles cancel channel by channel,
\begin{align}
 T_{q,\mathrm{red,loop}}^{(-1)\mu\nu}
 &=\frac{\mathcal V_{qq}}{4m_\mu}g_\perp^{\mu\nu}
 =-T_{q,\mathrm{red,GPD}}^{(-1)\mu\nu},
 \nonumber\\
 T_{g,\mathrm{red,loop}}^{(-1)\mu\nu}
 &=-T_{g,\mathrm{red,GPD}}^{(-1)\mu\nu},
 \qquad
 T_{g,\mathrm{red,GPD}}^{(-1)\mu\nu}\propto
 -g_\perp^{\mu\nu}\left(C_q^{(0)}\otimes V_{qg}^{[0]}\right)\; .
 \label{eq:nlo-pole-cancellation}
\end{align}
Here $C_q^{(0)}$ denotes the scalar reduced LO quark coefficient; the
proportionality sign suppresses common coupling and normalization factors.
The gluon matching uses the color-singlet and transverse-polarization
projector
\begin{equation}
 \frac{\delta^{ab}}{N_c^2-1}\frac{-g_\perp^{\alpha\beta}}{D-2}\ .
 \label{eq:gluon-projector}
\end{equation}
The finite functions $C_i^q$ and $C_i^g$ are collected in
Appendix~\ref{app:nlo-coefficients}.  The five coefficients are components of
the tensor basis in Eq.~\eqref{eq:nlo-factorization}, not independent CFFs.

For $x>0$, we use
$H^{q(+)}(x,\xi,t)=H^q(x,\xi,t)-H^q(-x,\xi,t)$ and
$H_{\mathrm{EM}}=(4H^{u(+)}+H^{d(+)}+H^{s(+)})/9$.  The gluon convention is
$H^g(x,0,0)=xg(x)$ with $H^g(-x,\xi,t)=H^g(x,\xi,t)$.  The quark and gluon
form factors are
\begin{align}
 \mathcal F_i^{(1)}={}&
 \frac{4C_F\as(\mu_R)}{\pi}
 \int_0^1\dd x\,H_{\mathrm{EM}}(x,\xi,t;\mu_F)
 C_i^q(x,\rho,\xi,Q^2;\mu_F,\mu_R),
 \nonumber\\
 \Gcal_i^{(1)}={}&
 \frac{8\as(\mu_R)}{3\pi}
 \int_{-1}^{1}\dd x\,H^g(x,\xi,t;\mu_F)
 C_i^g(x,\rho,\xi,Q^2;\mu_F),
 \qquad i=T,1,\ldots,4.
 \label{eq:nlo-five-cffs}
\end{align}
The prefactors restore the color, flavor, and gluon-polarization
normalizations omitted from the local hard functions.  The one-loop amplitude
is
\begin{equation}
 \mathcal M^{(1)}=
 \frac{e^4f_{\Dm}}{4Q^2}
 \bar u(k')\gamma_\nu u(k)\,\eps_\mu^*(P,\lambda_V)
 \bar u(p')\slashed{n}_{\mathrm{LC}}u(p)
 \sum_{i\in\{T,1,\ldots,4\}}
 (\mathcal F_i^{(1)}+\Gcal_i^{(1)})\tau_i^{\mu\nu}.
 \label{eq:nlo-amplitude}
\end{equation}
At the auxiliary collinear point, the reconstructed quark and gluon tensors
satisfy both electromagnetic Ward identities with respect to $q$ and $P_0$.
For comparison with Ref.~\cite{Pire:2011st}, we use
 $F_{\mathrm{PSW}}^g(x,\xi)=H^g(x,\xi)/x$ and
 $\mathcal Q_{\mathrm{PSW}}^2=Q^2/2$.  After this conversion, the contracted
transverse-helicity coefficients reproduce the known one-loop DVCS limits.
The comparison is made only after the full tensor is reconstructed; no
equality is implied for the five basis components separately.

\section{Numerical Results}
\label{sec:numerical-results}

\subsection{Numerical setup}
\label{sec:numerics}

The reference configuration consists of $18\,\GeV$ electrons and
$275\,\GeV$ protons.  The additional facility benchmarks in
Table~\ref{tab:facility-yields} follow
Refs.~\cite{AbdulKhalek:2022eic,Anderle:2021eicc,Baltzell:2020qoo,
Agostini:2021lhec}.  The fiducial region is
\begin{equation}
 Q^2>1\,\GeV^2,\qquad W>3\,\GeV,\qquad |t|<1\,\GeV^2,
 \label{eq:cuts}
\end{equation}
and the lower virtuality threshold is varied over
$1\,\GeV^2 \leq Q_{\min}^2\leq 100\,\GeV^2$.  Unless stated otherwise,
$\mu_F=\mu_R=\sqrt{Q^2+M_{\Dm}^2}$.  We take
$m_\mu=0.105658\,\GeV$, $M_{\Dm}=2m_\mu$, $m_p=0.938272\,\GeV$, and
$\alpha^{-1}=137.036$~\cite{ParticleDataGroup:2024cfk}, and use two-loop
running of $\as$ with heavy-flavor threshold matching.  The elastic proton
current is evaluated with
$G_E(t)=(1-t/0.71\,\GeV^2)^{-2}$ and $G_M(t)=\mu_pG_E(t)$, with
$\mu_p=2.792847$~\cite{Kelly:2004hm}.  Defining
$\tau_p=-t/(4m_p^2)$, the Dirac and Pauli form factors are
\begin{equation}
 F_1(t)=\frac{G_E(t)+\tau_p G_M(t)}{1+\tau_p},
 \qquad
 F_2(t)=\frac{G_M(t)-G_E(t)}{1+\tau_p}.
 \label{eq:elastic-form-factors}
\end{equation}

The model dependence is assessed with seven CFF inputs evaluated using a common hard coefficient, bound-state normalization, phase space, and set of cuts.  
VGG, GK19-H, and MMS13 use double-distribution constructions, and their numerical implementations follow the corresponding modules in \textsc{PARTONS}~\cite{Berthou:2015oaw}. 
They differ in their forward valence and sea inputs, profile functions, treatment
of the $D$ term, and correlated $t$ dependence
~\cite{Vanderhaeghen:1999xj,Kroll:2012sm,Mezrag:2013mya}.  The MMS13 input used here combines the MMS
valence sector with the GK sea sector, following the DDVCS implementation of
Ref.~\cite{Deja:2023ahc}.

EKM10 and EKM15 employ conformal Mellin--Barnes representations constrained
mainly by DVCS data~\cite{Kumericki:2015lhb}.  Their continuation from the DVCS line $\rho=\xi$ to
$\rho\neq\xi$ is therefore model dependent, especially at low $Q^2$.

The two FMS inputs start from the NLO MSTW2008 forward
distributions~\cite{Martin:2009iq}.  They differ in the construction of their
skewness dependence: one uses skewed evolution with an ERBL completion, while
the other uses the Shuvaev transform~\cite{Freund:2002qf,Shuvaev:1999ce}.
The CFF tables used here have no intrinsic $t$ dependence, so only these two inputs are supplemented by the amplitude-level profile
\begin{align}
 \mathcal M_{\mathrm{DDVCS}}(t)&=\mathcal M_{\mathrm{DDVCS}}(t_0)
 \exp\!\left[\frac{B_{\mathrm{H1}}(Q^2)}{2}(t-t_0)\right],
 \qquad t_0=t_{\min},
 \nonumber\\
 B_{\mathrm{H1}}(Q^2)&=A_t\left[1-b_t\ln\!\left(\frac{Q^2}{Q_0^2}\right)\right].
 \label{eq:empirical-t-slope}
\end{align}
The H1 fit gives $A_t=(6.98\pm0.54)\,\GeV^{-2}$,
$b_t=0.12\pm0.03$, and $Q_0^2=2\,\GeV^2$ for
$6.5<Q^2<80\,\GeV^2$~\cite{H1:2007vrx,H1:2009wnw}.  We use the central
values and freeze the slope at the endpoints outside this interval.  Models
with native $t$ dependence are not multiplied by an additional slope.
Consequently, the comparison includes differences in both skewness and
momentum-transfer dependence.

\subsection{Cross sections and distributions}
\label{sec:results}

\begin{figure}[!htbp]
\centering
\includegraphics[width=0.78\textwidth]{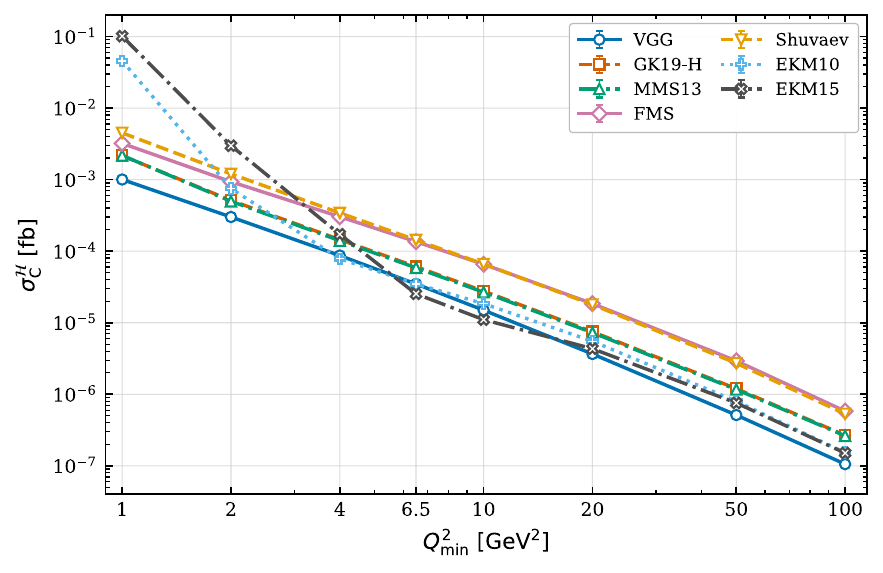}
\caption{Integrated $\Hcal$-only DDVCS cross section as a function of the
lower virtuality cut for the seven GPD/CFF inputs.}
\label{fig:model-scan}
\end{figure}

We first compare the $\Hcal$-only result, setting
$\Ecal=\Htcal=\Etcal=0$.  Fig.~\ref{fig:model-scan} shows the integrated
DDVCS cross section as the lower virtuality cut is raised.  The lowest-cut
sample maximizes the rate but is also the most sensitive to perturbative and
power corrections~\cite{Chen:2019pec,Flett:2021ghh}.
At $Q_{\min}^2=1\,\GeV^2$, the double-distribution and forward-PDF
constructions span $1.00\times10^{-3}$--$4.50\times10^{-3}\,\fb$, whereas
EKM10 and EKM15 give $4.57\times10^{-2}$ and
$1.01\times10^{-1}\,\fb$, respectively.  This order-of-magnitude model spread is
localized in the low-$Q^2$ continuation away from the DVCS line.  It contracts
to $7.83\times10^{-5}$--$3.43\times10^{-4}\,\fb$ at
$Q_{\min}^2=4\,\GeV^2$ and to
$(1.06$--$5.82)\times10^{-7}\,\fb$ at $100\,\GeV^2$.

\begin{figure}[!htbp]
\centering
\includegraphics[width=0.84\textwidth]{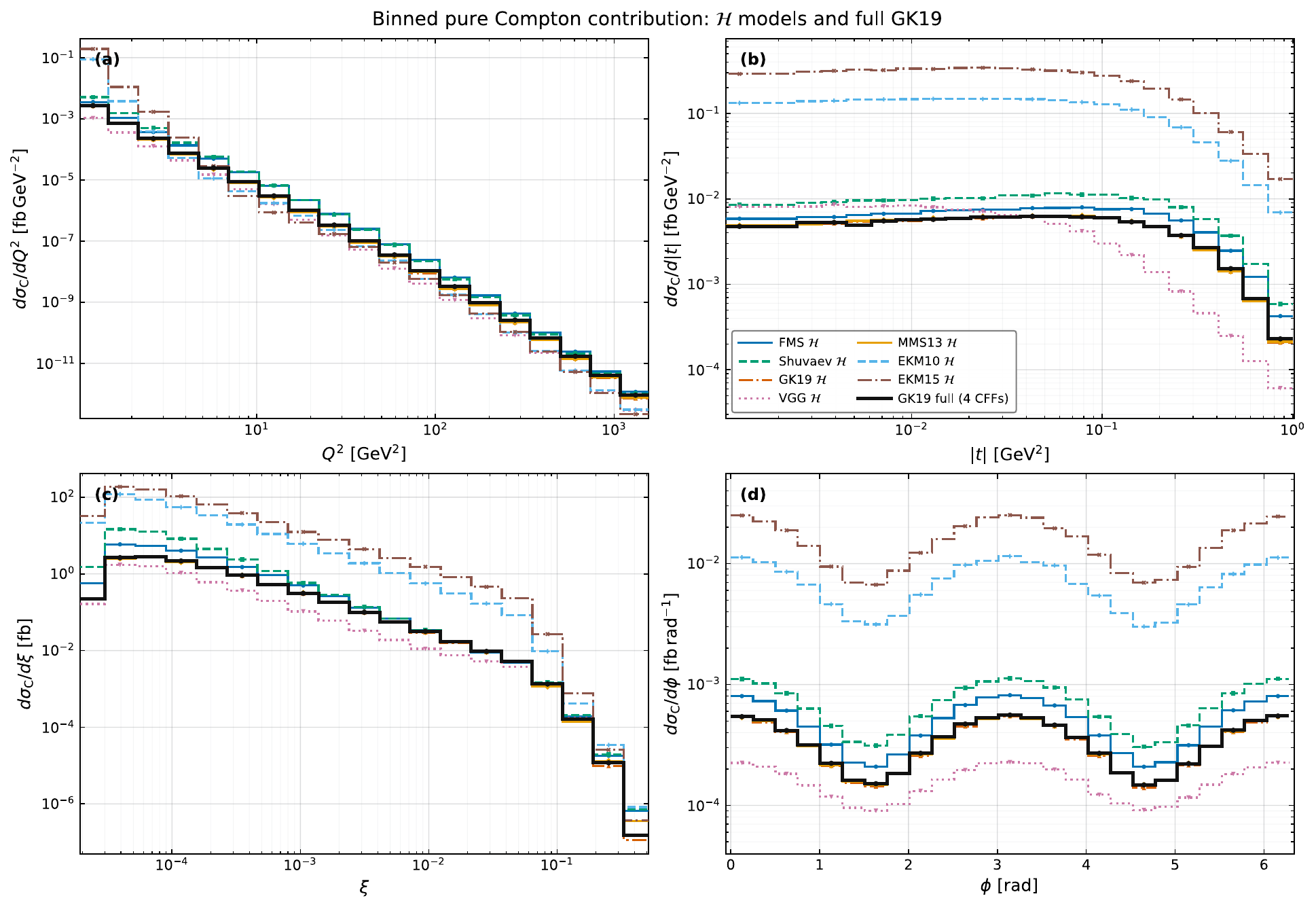}
\caption{LO DDVCS distributions in $Q^2$, $|t|$, $\xi$, and the
lepton--hadron azimuthal angle.  The curves compare the seven $\Hcal$ inputs
with the four-CFF GK19 result; BH production and BH--DDVCS interference are
not included.}
\label{fig:lo-distributions}
\end{figure}

Fig.~\ref{fig:lo-distributions} shows that DDVCS production is concentrated
at low $Q^2$, small $|t|$, and small skewness.  Both FMS inputs use the common
profile in Eq.~\eqref{eq:empirical-t-slope}; the other models retain their
native momentum-transfer dependence.

The total unpolarized rate is controlled by the BH contribution throughout
the virtuality range shown in Fig.~\ref{fig:bh-and-nonh}(a).  With all four
GK19 CFFs, $\sigma_{\mathrm{BH}}/\sigma_{\mathrm{DDVCS}}=16.8$, $108$, and $673$ for
$Q_{\min}^2=1$, $10$, and $100\,\GeV^2$, respectively.  The DDVCS fraction
therefore decreases rapidly as the virtuality cut is raised.
Fig.~\ref{fig:bh-and-nonh}(b) shows the relative change
$\sigma_{\mathrm{DDVCS}}^{4\,\mathrm{CFF}}/
\sigma_{\mathrm{DDVCS}}^{\Hcal}-1$ induced by including $\Ecal$, $\Htcal$,
and $\Etcal$ in addition to $\Hcal$.  The change is $2.52\%$, $8.41\%$, and
$14.5\%$ for $Q_{\min}^2=1$, $10$, and $100\,\GeV^2$, respectively.  Its
increasing relative size does not correspond to a larger cross section:
$\sigma_{\mathrm{DDVCS}}^{4\,\mathrm{CFF}}$ decreases from
$2.22\times10^{-3}\,\fb$ at $Q_{\min}^2=1\,\GeV^2$ to
$3.07\times10^{-7}\,\fb$ at $100\,\GeV^2$.

\par\addvspace{\intextsep}
\noindent\begin{minipage}{\textwidth}
\centering
\begin{minipage}[t]{0.40\textwidth}
\centering
\includegraphics[width=\linewidth]{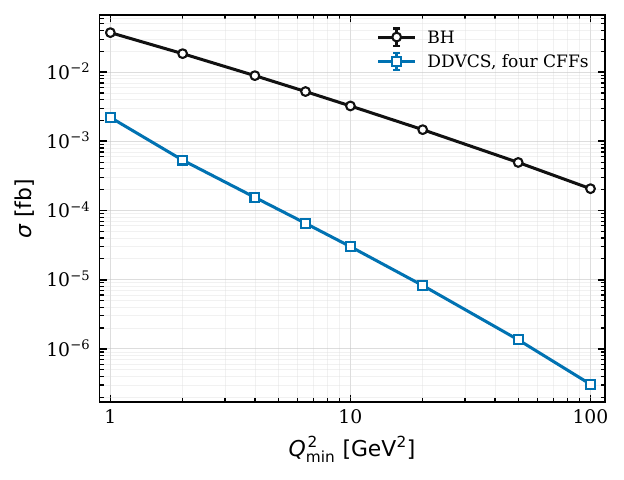}\\[0.5ex]
\textbf{(a)}
\end{minipage}\hfill
\begin{minipage}[t]{0.40\textwidth}
\centering
\includegraphics[width=\linewidth]{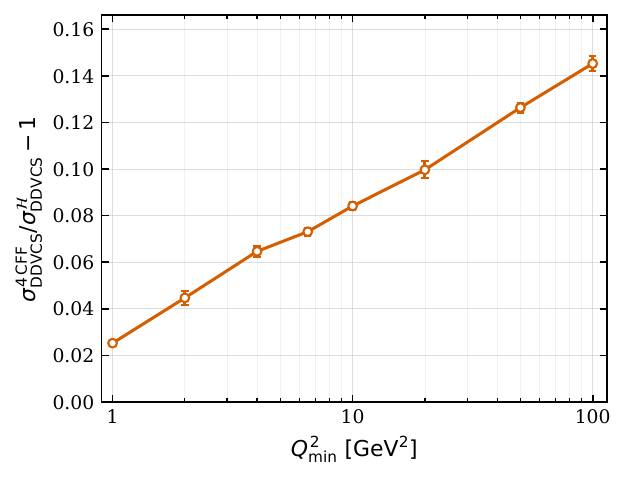}\\[0.5ex]
\textbf{(b)}
\end{minipage}
\captionof{figure}{(a) Integrated BH and four-CFF GK19 DDVCS cross sections
as functions of $Q_{\min}^2$.  (b) Relative change
$\sigma_{\mathrm{DDVCS}}^{4\,\mathrm{CFF}}/
\sigma_{\mathrm{DDVCS}}^{\Hcal}-1$ induced by including $\Ecal$, $\Htcal$,
and $\Etcal$.}
\label{fig:bh-and-nonh}
\end{minipage}
\par\addvspace{\intextsep}

The interference remains linear in the CFFs.  As in continuum DDVCS, charge,
spin, and azimuthal combinations can isolate harmonics proportional to their
real or imaginary parts~\cite{Belitsky:2003fj,Deja:2023ahc}.  Because BH dominates the unpolarized cross section,
interference observables provide more direct access to the CFF dependence than
the total rate.

We next examine the NLO QCD correction to the vector-$\Hcal$ DDVCS channel.
Table~\ref{tab:nlo-summary} separates the quark and gluon interference terms
for the FMS input.  Defining
$K_{\mathrm{DDVCS}}=\sigma_{\mathrm{DDVCS}}^{\mathrm{NLO}}/
\sigma_{\mathrm{DDVCS}}^{\mathrm{LO}}$ and
$K_{ep}=\sigma_{ep}^{\mathrm{NLO}}/\sigma_{ep}^{\mathrm{LO}}$, the lowest-cut Born
DDVCS rate is $3.208\times10^{-3}\,\fb$.  The quark and gluon terms contribute
$-4.948\times10^{-4}\,\fb$ and $-2.438\times10^{-3}\,\fb$, leaving
$2.752\times10^{-4}\,\fb$ at NLO.

\par\addvspace{\intextsep}
\noindent\begin{minipage}{\textwidth}
\centering
\captionof{table}{Integrated NLO quark and gluon corrections to the
vector-$\Hcal$ DDVCS channel for the FMS input.}
\label{tab:nlo-summary}
{\small
\setlength{\tabcolsep}{4.2pt}
\begin{ruledtabular}
\begin{tabular}{cccccc}
$Q_{\min}^2\ [\GeV^2]$ & $\sigma_{\mathrm{DDVCS}}^{\mathrm{LO}}\ [\fb]$ &
$\delta_q/\sigma_{\mathrm{DDVCS}}^{\mathrm{LO}}$ &
$\delta_g/\sigma_{\mathrm{DDVCS}}^{\mathrm{LO}}$ &
$K_{\mathrm{DDVCS}}$ & $K_{ep}$ \\
\hline
$1$   & $3.208\times10^{-3}$ & $-15.4\%$ & $-76.0\%$ & $0.0858$ & $0.9278$ \\
$2$   & $9.353\times10^{-4}$ & $-23.3\%$ & $+84.5\%$ & $1.612$ & $1.0293$ \\
$4$   & $3.030\times10^{-4}$ & $-22.3\%$ & $+141\%$ & $2.191$ & $1.0392$ \\
$6.5$ & $1.358\times10^{-4}$ & $-20.5\%$ & $+151\%$ & $2.304$ & $1.0330$ \\
$10$  & $6.557\times10^{-5}$ & $-19.0\%$ & $+150\%$ & $2.311$ & $1.0260$ \\
\end{tabular}
\end{ruledtabular}
}
\end{minipage}
\par\addvspace{\intextsep}

The gluon contribution is negative in the lowest-cut sample but becomes
positive already at $Q_{\min}^2=2\,\GeV^2$, reaching about
$+140\%$--$+150\%$ for $Q_{\min}^2\geq4\,\GeV^2$.  The sign and magnitude
refer to the full reconstructed gluon tensor and its interference with the LO
amplitude.  The negative lowest-cut correction is not caused by a negative gluon GPD, since
$H^g(x,0,0)=xg(x)$ in the adopted convention.  This behavior is qualitatively
consistent with the quark--gluon cancellation found in higher-order DVCS
studies~\cite{Moutarde:2013nlo,Braun:2022nnlo}, although the present DDVCS
trajectory, GPD input, and phase-space integration are different.  The large
$K$ factors do not establish perturbative convergence.  Existing NNLO DVCS
calculations show that the two-loop gluon term can strengthen the cancellation,
and the two-loop DDVCS coefficient also gives sizable corrections in the
examples of Ref.~\cite{Braun:2024jns}.

Fig.~\ref{fig:nlo-distributions} shows the virtuality dependence of the NLO
QCD correction in the differential distributions.  The negative lowest-$Q^2$
bin reflects destructive NLO interference in the low-scale region.

\par\addvspace{\intextsep}
\noindent\begin{minipage}{\textwidth}
\centering
\includegraphics[width=0.98\textwidth]{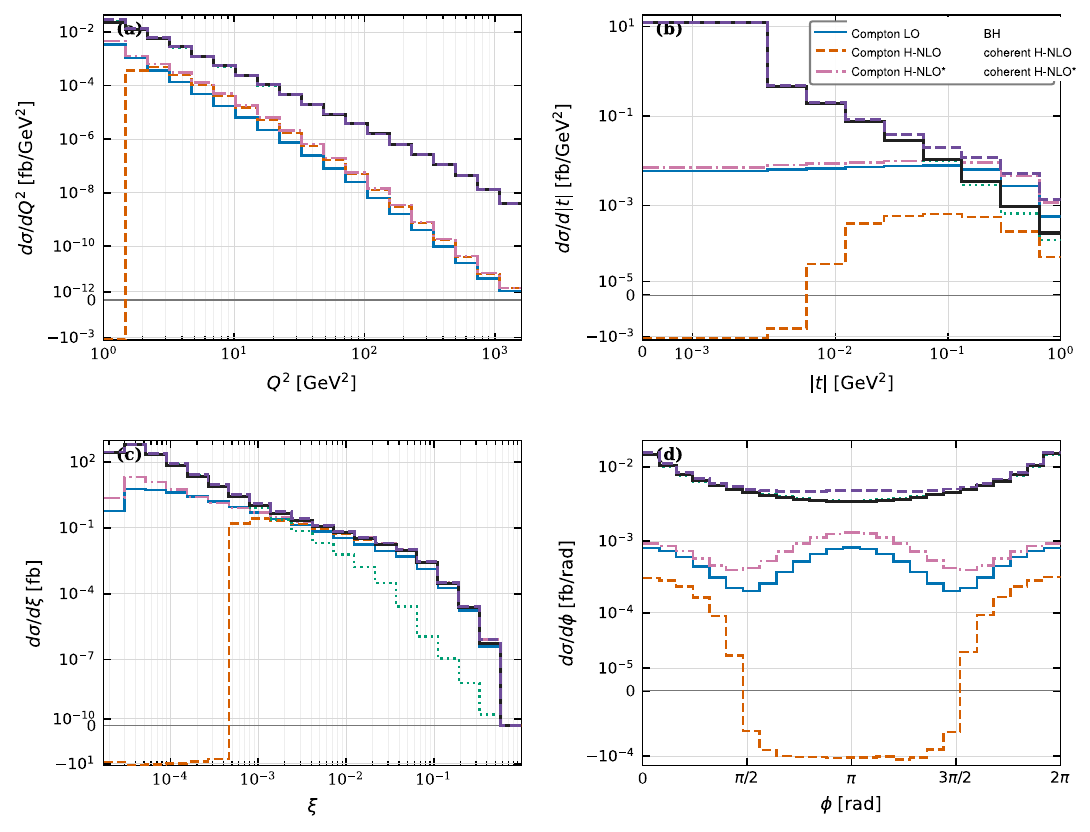}
\captionof{figure}{FMS differential distributions in (a) $Q^2$, (b) $|t|$,
(c) $\xi$, and (d) the lepton--hadron azimuthal angle.  The curves show Born
BH, LO DDVCS, NLO DDVCS and total results, together with the
corresponding NLO$^{*}$ quantities.  All panels use a symmetric
logarithmic vertical scale; the horizontal line marks zero, and negative bins
are displayed below it.}
\label{fig:nlo-distributions}
\end{minipage}
\par\addvspace{\intextsep}

The DDVCS $K$ factor rises from $0.0858$ at the lowest cut to about
$2.2$--$2.3$ for $Q_{\min}^2\geq4\,\GeV^2$.  By contrast, $K_{ep}$ remains
within $8\%$ of unity because BH dominates the total rate.  The
NLO$^{*}$ result includes $|\mathcal M^{(1)}|^2$; for the lowest cut,
$\sigma_{\mathrm{DDVCS}}^{\mathrm{NLO}^{*}}/\sigma_{\mathrm{DDVCS}}^{\mathrm{LO}}=1.50$.

Finally, we apply the same Born BH framework to positronium, dimuonium, and
tauonium.  Table~\ref{tab:facility-yields} lists the resulting full-phase-space
cross sections and one-year yields.

\par\addvspace{\intextsep}
\noindent\begin{minipage}{\textwidth}
\centering
\captionof{table}{Born BH cross sections integrated over the full physical
phase space and one-year production counts for the ${}^3S_1$ leptonium states,
with $T_{\mathrm{yr}}=3.15\times10^7\,\mathrm{s}$.}
\label{tab:facility-yields}
{\scriptsize
\setlength{\tabcolsep}{2.4pt}
\begin{ruledtabular}
\resizebox{\textwidth}{!}{%
\begin{tabular}{lcccccccc}
Facility & Beam energy $[\GeV]$ & $\mathcal L_{\mathrm{inst}}$ &
\multicolumn{2}{c}{positronium} & \multicolumn{2}{c}{dimuonium} &
\multicolumn{2}{c}{tauonium} \\
\cline{4-5}\cline{6-7}\cline{8-9}
 & & $[\mathrm{cm}^{-2}\mathrm{s}^{-1}]$ &
$\sigma_{\mathrm{BH}}^{\mathrm{full}}\ [\fb]$ & $N_{1\,\mathrm{yr}}$ &
$\sigma_{\mathrm{BH}}^{\mathrm{full}}\ [\fb]$ & $N_{1\,\mathrm{yr}}$ &
$\sigma_{\mathrm{BH}}^{\mathrm{full}}\ [\fb]$ & $N_{1\,\mathrm{yr}}$ \\
\hline
EIC    & $18\times275$      & $1.0\times10^{34}$ & $1.71\times10^6$    & $5.40\times10^8$ & $0.586$          & $185$             & $1.20\times10^{-3}$  & $0.378$ \\
EIC    & $10\times275$      & $1.0\times10^{34}$ & $2.95\times10^5$    & $9.29\times10^7$ & $0.542$          & $171$             & $9.93\times10^{-4}$  & $0.313$ \\
EicC   & $3.5\times20$      & $2.0\times10^{33}$ & $5.32\times10^3$    & $3.36\times10^5$ & $0.260$          & $16.4$            & $1.21\times10^{-4}$  & $7.62\times10^{-3}$ \\
JLab   & $11$ (fixed target) & $1.0\times10^{35}$ & $4.08\times10^3$    & $1.29\times10^7$ & $0.0954$         & $301$             & $1.55\times10^{-10}$ & $4.88\times10^{-7}$ \\
LHeC   & $50\times7000$     & $2.3\times10^{34}$ & $6.06\times10^{11}$ & $4.39\times10^{14}$ & $1.26$           & $916$             & $2.84\times10^{-3}$  & $2.06$ \\
FCC-eh & $60\times50000$    & $1.5\times10^{34}$ & $3.81\times10^{14}$ & $1.80\times10^{17}$ & $2.65\times10^2$ & $1.25\times10^5$ & $3.70\times10^{-3}$  & $1.75$ \\
\end{tabular}
}
\end{ruledtabular}
}
\end{minipage}
\par\addvspace{\intextsep}

The Born BH rates display a pronounced lepton-mass hierarchy.  The large
positronium samples provide a high-statistics environment for precision QED
studies, while the two
EIC configurations give $171$--$185$ dimuonium states per year and JLab, the
LHeC, and FCC-eh give about $301$, $916$, and $1.25\times10^5$, respectively.
Tauonium remains below one state per year except at the LHeC and FCC-eh, where
the yield is about two.  For positronium, only the two electron-line BH
topologies are included; exchange or rearrangement amplitudes associated with
the identical electrons are omitted.

The large BH contribution supplies the dominant exclusive event sample but
dilutes the DDVCS component in the unpolarized normalization.  This pattern
motivates angular, beam-charge, and spin observables that isolate BH--DDVCS
interference.

\FloatBarrier

\section{Conclusions}
\label{sec:summary}

We have studied exclusive vector-leptonium electroproduction in
$ep$ collisions through the BH and DDVCS mechanisms, including their
interference and the NLO QCD correction to the dominant vector-$\Hcal$
channel, and compared the Born BH rates of positronium, dimuonium, and
tauonium.  The vector bound state is treated in NRQED, while the DDVCS tensor
is factorized at twist 2 into perturbative coefficient functions and GPDs,
whose convolutions define the CFFs.  Bethe--Heitler production dominates the
unpolarized rate, supporting dedicated dimuonium searches at the EIC and JLab
and offering stronger prospects at the LHeC and FCC-eh.  The much larger
positronium samples provide a high-statistics environment for precision QED
studies, whereas tauonium production remains strongly suppressed.

The DDVCS contribution is concentrated at low $Q^2$, small $|t|$, and small
skewness.  Its model dependence is largest in the low-$Q^2$ region and
decreases as the virtuality cut is raised.  The contributions from $\Ecal$,
$\Htcal$, and $\Etcal$ remain small compared with $\Hcal$, although their
relative effect increases with the cut.  Because BH--DDVCS interference is
linear in the CFFs, angular, charge, and spin observables provide greater
relative sensitivity to the GPD-dependent amplitude than the unpolarized
normalization alone.

The NLO QCD correction to the vector-$\Hcal$ DDVCS channel is large, with its
sign and magnitude controlled by the balance between the quark and gluon
terms.  For the FMS input, the gluon contribution produces a strong
suppression at the lowest cut and a sizable enhancement once the cut is
raised, indicating limited fixed-order stability at the adopted scales.  The
available NNLO DVCS and DDVCS results suggest that higher-order singlet
corrections may remain important for this process.

\vspace{1.0cm} {\bf Acknowledgments}

The authors thank Ms. Yuan-Hui Zhu for assistance in organizing the research group.  This work was supported in part by the National Key Research and Development Program of China under Contract No.~2025YFA1613900, by the National Natural Science Foundation of China (NSFC) under Grants No.~12475087, 12235008, and 12235001, and by the University of Chinese Academy of Sciences.
\newpage
\appendix

\section{Analytic one-loop coefficient functions}
\label{app:nlo-coefficients}

The reduced coefficient functions used in the main text are collected here.
The common electromagnetic, bound-state, flavor, color, and coupling
prefactors are restored only through Eq.~\eqref{eq:nlo-five-cffs}.

We set $\widehat\rho=\rho-i0$.  The logarithms used in both channels are
\begin{equation}
 L_0=\ln\frac{\widehat\rho^2-x^2}{\widehat\rho^2-\xi^2},
 \qquad
 L_x=\ln\frac{\widehat\rho-x}{\widehat\rho+x},
 \qquad
 L_\xi=\ln\frac{\widehat\rho-\xi}{\widehat\rho+\xi}.
 \label{eq:nlo-logs}
\end{equation}
\subsection{Quark coefficient}
\label{app:quark-coefficients}

The convolution entering the quark counterterm is
\begin{align}
 \mathcal V_{qq}(x,\xi,\widehat\rho)
 \equiv{}&\left[\frac{x}{x^2-\widehat\rho^2}\otimes
 V_{qq}^{[0]}\right](x,\xi)
 \nonumber\\
 ={}&\frac{1}{2\xi(x^2-\xi^2)(x^2-\widehat\rho^2)}
 \Big[3x\xi(x^2-\xi^2)
 +\widehat\rho x(x^2-\widehat\rho^2)L_\xi
 \nonumber\\
 &\quad+\widehat\rho\xi(x^2+\widehat\rho^2-2\xi^2)L_x
 +x\xi(x^2+\widehat\rho^2-2\xi^2)L_0\Big].
 \label{eq:vqq-closed}
\end{align}
The transverse quark coefficient is
\begin{equation}
 C_T^q=\frac{[3+\ln(\mu_R^2/Q^2)]\mathcal N_R+\mathcal N_F}
 {m_\mu(x^2-\xi^2)(x^2-\widehat\rho^2)}
 +\ln\frac{\mu_F^2}{\mu_R^2}\frac{\mathcal V_{qq}}{4m_\mu},
 \label{eq:ctq-result}
\end{equation}
where
\begin{subequations}
\label{eq:nlo-quark-numerators}
\begin{align}
 \mathcal N_R={}&\frac18\Bigg[
 3x(x^2-\xi^2)
 +x(x^2+\widehat\rho^2-2\xi^2)L_0
 +\widehat\rho(x^2+\widehat\rho^2-2\xi^2)L_x
 +\frac{x\widehat\rho(x^2-\widehat\rho^2)}{\xi}L_\xi
 \Bigg],
 \\
 \mathcal N_F={}&-\frac{9x}{16}(x^2-\xi^2)L_0
 -\frac{x}{32}(x^2+\widehat\rho^2-2\xi^2)(L_0^2+L_x^2)
 -\frac{3\widehat\rho}{16}(x^2-3\xi^2+2\widehat\rho^2)L_x
 \nonumber\\
 &-\frac{x^2+\widehat\rho^2-2\xi^2}{16}
 (\widehat\rho L_0L_x+xL_0L_\xi+\widehat\rho L_xL_\xi)
 \nonumber\\
 &+\frac{3x}{16\xi}
 [\xi^3+2\widehat\rho^3-x^2(\xi+2\widehat\rho)]L_\xi
 \nonumber\\
 &+\frac{x}{32\xi}
 [-2\xi^3+x^2(\xi-2\widehat\rho)
 +\xi\widehat\rho^2+2\widehat\rho^3]L_\xi^2.
\end{align}
\end{subequations}
Writing $\mathcal L=L_x-(x/\xi)L_\xi$, the remaining quark coefficients are
\begin{align}
 C_1^q={}&\frac{m_\mu\mathcal L}
 {4(x^2-\xi^2)(\xi-\widehat\rho)},
 &
 C_2^q={}&\frac{(\xi+\widehat\rho)\mathcal L}
 {8m_\mu(x^2-\xi^2)},
 \nonumber\\
 C_3^q={}&-\frac{(\xi-\widehat\rho)\mathcal L}
 {8m_\mu(x^2-\xi^2)},
 &
 C_4^q={}&-\frac{(\xi-\widehat\rho)^2(\xi+\widehat\rho)\mathcal L}
 {16m_\mu^3(x^2-\xi^2)}.
 \label{eq:quark-nontransverse-coefficients}
\end{align}

\subsection{Gluon coefficient}
\label{app:gluon-coefficients}

For compactness, define
\begin{align}
 \mathcal D_g={}&(x^2-\xi^2)^2,
 \qquad
 \Delta_F=3+\ln\frac{\mu_F^2}{Q^2},
 \nonumber\\
 \mathcal A_g={}&2\xi\widehat\rho L_0-2x\xi L_x
 +(x^2+\xi^2)L_\xi,
 \nonumber\\
 \mathcal B_g={}&\xi(x^2-\xi^2+2\widehat\rho^2)L_0
 -2x\xi\widehat\rho L_x
 +\widehat\rho(x^2+\xi^2)L_\xi,
 \label{eq:gluon-auxiliary-functions}
\end{align}
\begin{align}
 \mathcal R_g={}&
 \xi(-x^2+\xi^2-2\widehat\rho^2)(L_0^2+L_x^2)
 \nonumber\\
 &+L_\xi\Big[-8x^2\widehat\rho
 +\big(x^2(\xi-2\widehat\rho)
 -\xi(\xi^2+2\xi\widehat\rho-2\widehat\rho^2)\big)L_\xi\Big]
 \nonumber\\
 &+4x\xi\widehat\rho L_x(2+L_\xi)
 \nonumber\\
 &+2\xi L_0\Big[-4(x^2-\xi^2+\widehat\rho^2)
 +2x\widehat\rho L_x
 +(-x^2+\xi^2-2\widehat\rho^2)L_\xi\Big].
 \label{eq:gluon-r-function}
\end{align}
The finite gluon coefficients are
\begin{align}
 C_T^g={}&\frac{(\xi-\widehat\rho)
 [4\Delta_F\mathcal B_g+\mathcal R_g]}
 {128m_\mu^3\xi\mathcal D_g(\xi+\widehat\rho)},
 \label{eq:ctg-result}\\
 C_1^g={}&\frac{\mathcal A_g}
 {16m_\mu\xi\mathcal D_g(\xi+\widehat\rho)},
 &
 C_2^g={}&\frac{(\xi-\widehat\rho)\mathcal A_g}
 {32m_\mu^3\xi\mathcal D_g},
 \nonumber\\
 C_3^g={}&-\frac{(\xi-\widehat\rho)^2\mathcal A_g}
 {32m_\mu^3\xi\mathcal D_g(\xi+\widehat\rho)},
 &
 C_4^g={}&-\frac{(\xi-\widehat\rho)^3\mathcal A_g}
 {64m_\mu^5\xi\mathcal D_g}.
 \label{eq:gluon-nontransverse-coefficients}
\end{align}
The $i0$ prescription in $\widehat\rho$ fixes the branches in both the DGLAP
and ERBL regions.  The apparent singularities of the five-tensor decomposition
cancel in physical helicity amplitudes.

\end{document}